\documentclass[prl,twocolumn]{revtex4}
\usepackage[T1]{fontenc}
\usepackage[cp1250]{inputenc}
\usepackage{graphicx}

\begin{document}
\title{\bf Localization of maximal entropy random walk}
\author{Z. Burda$^1$}\email{zdzislaw.burda@uj.edu.pl}
\author{J. Duda$^1$}
\author{J.M. Luck$^2$}
\author{B. Waclaw$^3$}

\affiliation{
$^1$\mbox{Marian Smoluchowski Institute of Physics,
Jagellonian University, Reymonta 4, 30-059 Krak\'ow, Poland}
$^2$\mbox{Institut de Physique Th\'eorique, CEA IPhT and CNRS URA 2306, CEA Saclay, 91191 Gif-sur-Yvette cedex, France}
$^3$\mbox{Institut f\"ur Theoretische Physik, Universit\"at Leipzig,
Postfach 100\,920, 04009 Leipzig, Germany} 
}

\begin{abstract}
We define a new class of random walk processes which maximize entropy. This maximal entropy random walk is equivalent to generic random walk if it takes place on a regular lattice, but it is not if the underlying lattice is irregular. In particular, we consider a lattice with weak dilution. We show that the stationary probability of finding a particle performing maximal entropy random walk localizes in the largest nearly spherical region of the lattice which is free of defects. This localization phenomenon, which is purely classical in nature, is explained in terms of the Lifshitz states of a certain random operator.
\end{abstract}

\maketitle

Since the seminal papers by Einstein~\cite{ref:einstein}
and Smoluchowski~\cite{ref:smol} which formulated the theory of
Brownian motion and diffusive processes, a discrete-time realization
of these processes, random walk (RW), has continuously attracted
attention. Random walk has been discussed in thousands
of scientific papers and textbooks in statistical physics,
economics, biophysics, engineering, particle physics, etc.,
and is still an active research area (see e.g.~\cite{ref:rw}).
Mathematically speaking, random walk is a Markov chain which
describes the trajectory of a particle (random walker) taking
successive random steps. In the simplest variant, random walk
on a lattice, at each time step the particle chooses
at random one of the adjacent nodes and jumps to it.
In the continuum limit, the probability density of finding the particle at
a given position obeys the diffusion equation.
When the lattice is regular, it is easy to show that all
trajectories (sequences of nodes visited by the particle) of a given length
between two given points of the lattice are equiprobable,
and thus have maximal entropy. Therefore, drawing an analogy
with the path-integral formalism~\cite{ref:feynman},
trajectories are weighted only by their length,
which plays the role of the action in the absence of potential energy.

In this Letter we ask what changes if one takes the above statement
as a {\it definition} of RW.
In other words, we define random walk not by local hopping rules but by
the requirement that RW trajectories maximize entropy.
We shall see that, if the lattice is not regular,
this new definition leads to a dramatic change in the behavior of RW.
Let us summarize our main results. First, we define the
maximal entropy random walk (MERW) and show that it indeed maximizes the entropy
of trajectories, in contrast to generic random walk (GRW), which
has smaller entropy. Second, we discuss a surprising effect of
localization of MERW trajectories in the presence of weak disorder.
This is a purely classical example of the Lifshitz phenomenon~\cite{ref:lif}.
Some kind of localization has been observed
before in RW on networks with a broad distribution of nodes degrees
\cite{ref:noh}, but for MERW the effect is completely different in nature,
since it can be triggered by any small amount of inhomogeneity.

To begin, let us consider quite generally
a particle hopping randomly from node to node
on a given finite, connected graph.
The graph is defined by a symmetric adjacency matrix $A$,
with elements
$A_{ij}=1$ if $i$ and $j$ are neighboring nodes and $A_{ij}=0$ otherwise.
The hopping is a local Markov process: the particle
which arrives at some moment to node $i$ will hop to a neighboring
node $j$ with probability
$P_{ij}$, independently of the past history.
The elements of the transition matrix are $P_{ij}=0$ if $A_{ij}=0$,
that is if nodes $i,j$ are not linked,
and for each $i$ one has $\sum_j P_{ij} = 1$.

The main quantity of interest is the probability, $\pi_i(t)$,
of finding the particle at node $i$ at time $t$. One can
calculate it recursively, applying the Markov
property:
\begin{equation}
\pi_i(t+1) = \sum_j \pi_j(t) P_{ji}.
\label{markov1}
\end{equation}
Using spectral properties of the matrix $P_{ij}$, one can show
that $\pi_i(t)$ reaches for $t \rightarrow \infty$ a unique
stationary state $\pi_i^*$ obeying the following eigenequation:
\begin{equation}
\pi_i^* = \sum_j \pi_j^* P_{ji}.
\label{stationary}
\end{equation}
For GRW, $P_{ij} = A_{ij}/k_i$, where $k_i= \sum_j A_{ij}$ is the
number of neighbors of node $i$ (node degree). This means that
the particle hops to an adjacent
node with the same probability for all neighbors.
The stationary distribution of GRW reads
\begin{equation}
\pi_i^* = \frac{k_i}{\sum_j k_j}.
\label{stat1}
\end{equation}

Another quantity of interest, especially important from the point
of view of RW entropy, is the probability
$P(\gamma^{(t)}_{i_0 i_t})$ of generating a trajectory
$\gamma_{i_0 i_t}^{(t)}$ of length $t$, passing through nodes
$(i_0,i_1,\ldots,i_{t-1},i_t)$:
\begin{equation}
P(\gamma^{(t)}_{i_0 i_t}) = P_{i_0 i_1} P_{i_1i_2} \ldots P_{i_{t-1} i_t}.
\label{path_prob}
\end{equation}
In general $P(\gamma^{(t)}_{i_0 i_t})$ depends on all
nodes on the trajectory. For GRW we have
\begin{equation}
P(\gamma^{(t)}_{i_0 i_t}) = \frac{1}{k_{i_0} k_{i_1} \ldots k_{i_{t-1}}},
\label{path_GRW}
\end{equation}
and we see that the trajectories are not equiprobable.
An exception is GRW on a $k$-regular graph, whose nodes have
identical degrees, as for instance on a regular lattice.
In general, however, trajectories produced by GRW
are not maximally random. As we will see below, there exists,
though, a choice of $P_{ij}$ such that
all trajectories of given length~$t$ and given endpoints are equiprobable.
This choice corresponds to MERW.

Let us now present the explicit construction of MERW.
Let $\psi_i$ be the normalized eigenvector,
$\sum_i \psi_i^2 =1$, corresponding to the maximal eigenvalue $\lambda$
of the adjacency matrix $A_{ij}$:
\begin{equation}
\sum_j A_{ij} \psi_j = \lambda \psi_i.
\label{adj_max}
\end{equation}
The eigenvalue
$\lambda$ is clearly in the range $k_{\rm min} \le \lambda \le k_{\rm max}$,
where $k_{\rm min}$ and $k_{\rm max}$
are the maximal and minimal node degrees of the graph, respectively.
The Frobenius-Perron theorem tells us that
the eigenvector has all elements of the same sign, so that one can choose
$\psi_i>0$. Let us use this eigenvector to define the
following transition matrix:
\begin{equation}
P_{ij} = \frac{A_{ij}}{\lambda} \frac{\psi_j}{\psi_i}.
\label{merw}
\end{equation}
By construction, the entries $P_{ij}$ are positive if $i$ and $j$ are
neighboring nodes. They are also properly normalized:
$\sum_j P_{ij} = 1$. A similar construction has been recently proposed
in the context of optimal information coding~\cite{ref:jd}.
The weight (\ref{path_prob}) is now independent of intermediate nodes:
\begin{equation}
P(\gamma^{(t)}_{i_0 i_t}) = \frac{1}{\lambda^{t}} \frac{\psi_{i_t}}{\psi_{i_0}}, \label{equiprob}
\end{equation}
and thus all trajectories having length $t$ and given endpoints $i_0$
and $i_t$ are equiprobable. For a closed trajectory, the probability
(\ref{equiprob}) depends only on its length $t$.
The stationary distribution of MERW is
\begin{equation}
\pi_i^* = \psi_i^2,
\label{stat_merw}
\end{equation}
which is easy to check by combining Eqs.~(\ref{merw}) and (\ref{stationary}).
It is a normalized probability: $\sum_i \pi_i^* = 1$,
and the detailed balance condition is fulfilled:
$\pi_i^* P_{ij} = \pi_j^* P_{ji}$.

We intuitively see that random trajectories generated by the transition
probabilities $(\ref{merw})$ are more random than those generated
by GRW since now the probability of a given random
path (\ref{equiprob}) is independent of intermediate nodes.
This statement can be quantified by comparing the entropy rates of the
corresponding Markov processes. Let $P(i_0,i_1, \ldots,i_t)$ be
the probability of a sequence $(i_0,i_1,\ldots,i_t)$ in the set
of all sequences of length~$t$ generated by the Markov chain. The
Shannon entropy in this set of sequences is:
\begin{equation}
S_t = - \sum_{i_0,i_1\dots i_t} P(i_0,\ldots,i_t) \ln P(i_0,\ldots,i_t).
\end{equation}
One can show~\cite{ref:sm}, using the Markov property of the chain:
$P(i_0,i_1,\ldots,i_t) = \pi_{i_0} P_{i_0i_1} \ldots P_{i_{t-1}i_t}$,
that for large $t$ the entropy $S_t$ increases at a fixed rate
\begin{equation}
s \equiv \lim_{t\rightarrow \infty} \frac{S_t}{t}
= -\sum_i \pi^*_i \sum_{j} P_{ij} \ln P_{ij},
\end{equation}
which is independent of the initial distribution $\pi_{i}$. For GRW,
with $P_{ij} =A_{ij}/k_i$ and $\pi^*_i$ from Eq.~(\ref{stat1}), we obtain
the entropy production rate
\begin{equation}
s_{\rm GRW} = \frac{\sum_i k_i \ln k_i}{\sum_i k_i},
\label{sgrw}
\end{equation}
while MERW, with transition rates (\ref{merw})
and the stationary distribution (\ref{stat_merw}), yields $s_{\rm MERW} = \ln \lambda.$
We now show that $s_{\rm MERW}$ is indeed the maximal entropy
rate which can be obtained for any stochastic process generating
trajectories on the graph.
The number of trajectories of length $t$ on the graph is $N_t = \sum_{i,j} (A^t)_{ij}$,
where $A^t$ is the $t$-th power of the adjacency matrix.
In the $t\rightarrow \infty$ limit we obtain the following asymptotic value:
\begin{equation}
s_{\rm MAX} = \lim_{t\rightarrow \infty} \frac{\ln N_t}{t} = \ln \lambda,
\end{equation}
which sets the upper limit for the entropy rate of such processes.
We see that $s_{\rm MERW}=s_{\rm MAX}$, so that MERW indeed maximizes
the entropy and the corresponding trajectories are maximally random.
As a byproduct we obtain a lower bound for
the largest eigenvalue of the adjacency matrix:
\begin{equation}
\ln \lambda \ge \frac{\sum_i k_i \ln k_i}{\sum_i k_i},
\end{equation}
which we have not found in the literature. For a $k$-regular graph,
$s_{\rm GRW}=s_{\rm MERW}=\ln k$. Similarly,
for a bipartite graph which has nodes of degree $k$ in one
partition and of degree $k'$ in the other one,
$s_{\rm GRW}=s_{\rm MERW}=\frac12\ln(kk')$.

\begin{figure}
	\includegraphics*[width=7cm]{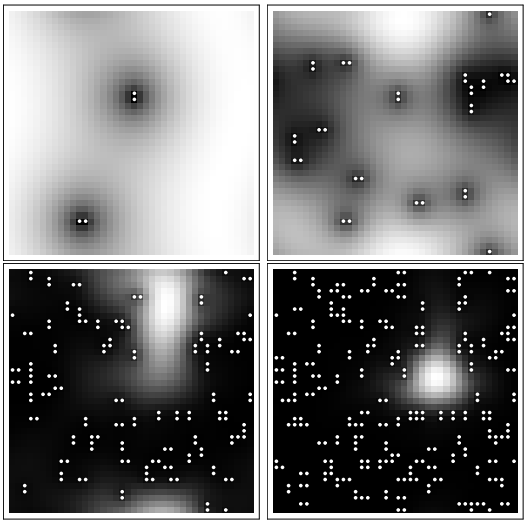}
	\caption{\label{fig1}Density plots of $\pi_i^*$ for a $40\times 40$
	square lattice with periodic boundary conditions,
	for the fractions $q=0.001,0.01,0.05,0.1$ of removed links.
	The nodes incident with removed links are marked with circles.
	Data are obtained by an exact diagonalization of the adjacency matrix.}
\end{figure}

As already mentioned, GRW and MERW are identical on a $k$-regular graph.
For example, GRW on a square lattice is maximally random. The question arises
how much the two types of random walk differ on a graph or lattice
with some irregularities.
For definiteness, imagine that we remove at random a small
fraction $q\ll1$ of non-adjacent links from an $L \times L$ square lattice
with periodic boundary conditions.
In this way we obtain a lattice with a weak disorder (dilution),
where most of the nodes are of degree $k=4$ and some of
degree $k=3$. The stationary distribution $\pi_i^*$ for
GRW is given by Eq.~(\ref{stat1}), so that the probability
of finding the particle after long time at a defective node
is equal to 3/4 of the probability at an intact one. The situation
looks completely different for MERW, as shown in Fig.~\ref{fig1},
presenting density plots of $\pi_i^*$ for different densities of defects,
obtained by diagonalizing
$A$ numerically and using Eq.~(\ref{stat_merw}).
For a very low density $q$ of defects,
the probability $\pi_i^*$ is smaller
in the neighborhood of defects, like in the GRW case.
However, if the number of defects increases, the stationary
distribution $\pi_i^*$ becomes localized in a nearly circular region.
We will indeed argue, using the Lifshitz argument~\cite{ref:lif},
that this localization phenomenon
is observed for any finite fraction of defects
provided the linear size $L$ of the system is large enough,
and that the radius of the localization region grows as $(\ln L)^{1/2}$.

\begin{figure}
	\includegraphics*[width=7cm]{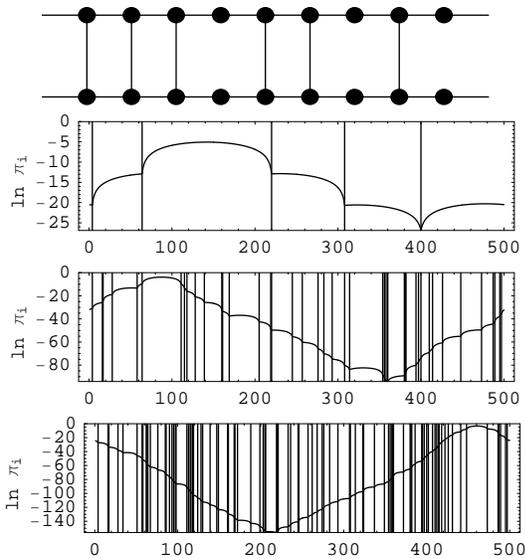}
	\caption{\label{fig2} Top: a ladder with randomly removed rungs.
	Bottom: stationary distributions $\pi_i^*$ on the ladder for $L=500$
	and various densities of defects $q=0.01,0.1,0.2$.
	Positions of defects are marked with vertical lines.}
\end{figure}

Let us start with a 1d example, in order to build up some
intuition. One cannot of course use a one-dimensional chain,
since removing a single link would disconnect it.
Instead, we shall consider, as a model example, a ladder graph
with periodic boundary conditions, with a fraction $q$ of
randomly removed rungs, as shown in Fig.~\ref{fig2}. In order to
define the transition probabilities (\ref{merw}) we have to solve
the eigenproblem of the adjacency matrix $A$.
Let $L$ be length of the ladder.
Taking into account the symmetry between both legs, we have:
\begin{equation}
\psi_{i+1} + \psi_{i-1} + r_i \psi_{i} = \lambda \psi_i,
\label{1d}
\end{equation}
where the index $i$ runs over the $L$ nodes in the lower leg of the ladder,
say, and $r_i=1$ if there is a rung at the position $i$, and $r_i=0$ otherwise.
Introducing the discrete Laplacian
$\Delta_{ij}=\delta_{i,j+1}+\delta_{i,j-1}-2\delta_{ij}$,
Eq.~(\ref{1d}) can be recast as
\begin{equation}
-(\Delta\psi)_i+v_i\psi_i=E\psi_i,
\label{schrodinger}
\end{equation}
where $E=3-\lambda$,
whereas $v_i=1-r_i$ form a random binary sequence with
a frequency of unities or defects ($v_i=1$) equal to $q$
and a frequency of zeros ($v_i=0$) equal to $p=1-q$.
Each sequence of sites without defects ($v_i=0$) is said to form a well.
Eq.~(\ref{schrodinger}) is formally identical
to the eigenvalue equation of the following trapping problem.
A particle performs a random walk in continuous time on the 1d lattice.
Defects act as static traps:
whenever the particle sits at site $i$,
it is annihilated at rate $v_i$ per unit time.
Trapping problems of this kind have been studied extensively~\cite{ref:trap}.
The asymptotic long-time fall-off of the survival probability
is known to be related to the so-called Lifshitz tail in the density of states
of Eq.~(\ref{schrodinger}) as $E\to0$.
In the present context, the Lifshitz argument~\cite{ref:lif}
predicts that the ground state of Eq.~(\ref{schrodinger})
is well approximated by that of the longest well, i.e.,
$-(\Delta\psi)_i = E_0 \psi_i$ ($i=1,\dots,w$),
with Dirichlet boundary conditions $\psi_0=\psi_{w+1}=0$,
where $w$ is the length of that well.
We obtain $\psi_i\sim\sin(i\pi/(w+1))$
and $E_0=2(1-\cos\pi/(w+1))\approx\pi^2/w^2$.
In the 1d situation~\cite{ref:jml},
this argument is known to essentially give an exact
description of the ground-state.

\begin{figure}
\includegraphics*[width=7cm]{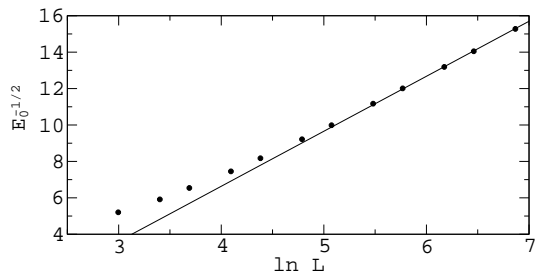}
\caption{\label{fig3} Ground-state energy $E_0$ of Eq.~(\ref{schrodinger})
on ladders versus $\ln L$ for $L=20,\dots,960$ and $q=0.1$.
The solid line shows the estimate $E_0^{-1/2} = \ln L /(\pi |\ln p|)+B$,
with $B$ fitted to the rightmost data point.}
\end{figure}

In the case of MERW, we therefore predict that the whole stationary probability
is asymptotically localized on the longest well,
i.e., the longest sequence without defects.
The Lifshitz picture is indeed a good approximation,
as one can see in Fig.~\ref{fig2}, showing the density $\pi_i^*$,
obtained by numerical diagonalization of $A$.
The length~$w$ of the longest well can be estimated as follows.
The mean number of unities in the sequence grows as $Lq$.
The mean number of those followed by one zero is $Lqp$,
by two zeros is $Lqp^2$, and so on, so that there are $Lq p^n$
wells of length $n$, i.e., consisting of $n$ zeros.
The length of the longest well is then given by $Lqp^w\sim1$.
Hence it grows logarithmically with the system size, $w\approx\ln L/|\ln p|$,
so that $E_0\approx(\pi|\ln p|/\ln L)^2$.
In Fig.~\ref{fig3} we show that the ground-state energy
$E_0$ obtained by numerically solving
Eq.~(\ref{schrodinger}), averaged over
binary disorder for $q=0.1$, agrees with the above estimate
for $L$ large enough.

The Lifshitz argument can be generalized to higher-dimensional
lattices~\cite{ref:mplif}.
The ground state of the discretized Schr\"odinger equation~(\ref{schrodinger})
is localized in the largest {\it Lifshitz sphere},
defined as the largest nearly spherical region of the lattice which is free of defects.
Taking again for definiteness the example of the square lattice,
the radius $R_{\rm max}$ of the largest Lifshitz disk
and the corresponding ground-state energy $E_0$ can be evaluated as follows.
The number of circular regions of radius $R$ with no defects
is of order $L^2 p^{2\pi R^2}$, as there are two links per node,
so that $R_{\rm max}\approx(\ln L/(\pi|\ln p|))^{1/2}$.
In the continuum limit, the ground state in the disk of radius
$R$ is given by $\psi(r)\sim J_0(jr/R)$, where $r$
is the distance from the center and $j\approx2.405$ is the
first zero of the Bessel function $J_0$.
We thus obtain $E_0\approx(j/R_{\rm max})^2\approx\pi j^2|\ln p|/\ln L$.
In higher dimension, skipping constants, the above estimates read
$R_{\rm max}\sim(\ln L/|\ln p|)^{1/d}$ and $E_0\sim(|\ln p|/\ln L)^{2/d}$.
Hence the stationary probability of MERW
on a $d$-dimensional lattice in the presence of any amount of disorder
is localized in the largest Lifshitz sphere,
whose {\it volume} grows asymptotically as $\ln L$.

The above picture allows one to address dynamical issues.
Imagine a random walker starting at a random site.
In the course of evolution it will find a moderately large region
free of defects,
a sort of local Lifshitz sphere, and spend some time there before it will make an excursion
to another, larger local Lifshitz sphere, etc.
The process will look very much like going through consecutive
metastable states before finally reaching the true ground state.
Metastable states are formed not because of energy barriers,
but because of entropy barriers~\cite{ref:fr},
as MERW favors regions where it can maximize entropy.
It is therefore tempting to consider MERW as a model of evolution
in a flat fitness landscape.

Let us close up with a comment on the connection
with the path-integral formalism~\cite{ref:feynman}.
In the simplest case of a free particle propagating
in curved space-time from $a$ to~$b$, the quantum amplitude is
\begin{equation}
K_{ab} = \sum_t \sum_{\gamma^{(t)}_{ab}} {\rm e}^{-S_{\rm E}},
\end{equation}
where the Euclidean action $S_{\rm E}$ is proportional to time~$t$.
One would naively expect that all the trajectories $\gamma_{ab}^{(t)}$
should be equiprobable.
We know, however, that the propagator $K_{ab}$ of a massless scalar field is
equal to the inverse of the graph Laplacian $\Delta_{ab} = \sqrt{k_a k_b} \delta_{ab} - A_{ab}$
which can be expressed as a sum over GRW and not MERW trajectories.
It would be interesting to check to what extent
the continuum theory would differ
if one constructed quantum amplitudes using MERW instead of GRW trajectories.

In conclusion, we have shown that GRW maximizes local entropy and MERW
maximizes global entropy of random trajectories. This little
change in the definition of random walk leads to a dramatic change
in the statistical properties of the system in the presence of a weak disorder.

We would like to thank P. Bialas and F. David for discussions.
This work was supported in part by the Marie Curie Actions Transfer of
Knowledge project ``COCOS'' -- grant No.~MTKD-CT-2004-517186,
the EC-RTN Network ``ENRAGE'' -- grant No.~MRTN-CT-2004-005616
and the Polish Ministry of Sciences and Information
Technologies -- grant 1P03B-04029 (2005-2008).

\end{document}